\documentstyle[12pt]{article}

\def\'#1{{\accent19\ifx #1i \i\else #1\fi}}

\def\be{\begin{equation}}
\def\ee{\end{equation}}
\def\bea{\begin{eqnarray}}
\def\eea{\end{eqnarray}}

\newbox\Ancha
\catcode`@=11
\newdimen\ex@
\title{Standard-model
particles and interactions  from field equations on spin 9+1
dimensional space}

\author{J. Besprosvany}

\date{Instituto de F\'{\i}sica, Universidad Nacional Aut\'onoma de M\'exico,
Apartado Postal 20-364, M\'exico 01000, D. F., M\'exico }

\begin{document}

\maketitle









\jot = 1.5ex
\def\baselinestretch{1.10}
\parskip 5pt plus 1pt

\begin{abstract}
We consider a Dirac equation set on an extended spin space that
contains fermion and boson solutions. At given dimension, it
determines the scalar symmetries.  The standard field equations
can be equivalently written  in terms of  such degrees of freedom,
and are similarly constrained. At 9+1 dimensions, the $SU(3)\times
SU(2)_L\times U(1)$ gauge groups emerge, as well as solution
representations with quantum numbers of  related gauge bosons,
leptons, quarks, Higgs-like particles and others as lepto-quarks.
Information on the coupling constants is also provided; e. g., for
the hypercharge $g^\prime=\frac{1}{2}\sqrt{\frac{3}{5}}\approx
.387$, at tree level.
\end{abstract}
\vskip 2cm \centerline{PACS: 11.10.Kk, 11.15.-q, 12.10.Dm,
12.60.-i.}

\vskip 1cm

\baselineskip 20pt\vfil\eject \noindent

The current theory of elementary particles, the standard model
(SM) is successful in describing their behavior, but it is
phenomenological. The origin  of  the interaction groups, the
particles' spectrum and representations, and parameters has
remained largely unexplained. Still, the generalization of
features of the model into larger structures with a unifying
principle has suggested connections among the observables. Thus,
additional dimensions in Kaluza-Klein theories are associated
with gauge interactions, and
 larger groups in grand-unified theories$\cite{unification}$ put some restrictions
on them. Spin is a physical manifestation of the
  fundamental representation of the Lorentz group
 and it is more so   in relation to
 space,  which uses the vector representation.
By setting Poincar\'e-invariant field equations on an extended
spin space$\cite{Jaime}$, described by a Clifford algebra, this
letter finds restrictions on and classifies the symmetries and the
representation solutions, at given dimension. They contain
spin-1/2 and boson solutions with a fixed scalar representation.
 A field theory can be
formulated in terms of such degrees of freedom. By constraining
only the dimension of the spin space on which the fields are
formulated, we derive  the interactions and representations of the
SM at 9+1 dimensions, the minimal space that contains them.







The Dirac equation
\begin{eqnarray}
\label {Jaimeq} \gamma_0( i \partial_\mu\gamma^\mu -M)\Psi ={ 0},
\end{eqnarray}
uses an extended spin space  when $\Psi$ represents a matrix
instead of, as traditionally, a  four-entries (column) spinor. Eq.
\ref{Jaimeq} contains four conditions over four spinors in a
$4\times 4$ matrix.
 There are, then, additional possible transformations and symmetry operations
 that further classify $\Psi$.
 The Dirac-operator transformation $( i \partial_\mu\gamma^\mu -M)\rightarrow
{U}(\it i
\partial_\mu\gamma^\mu -M) {U^{-1}}$
 induces
the left-hand side of the transformation
\begin{eqnarray}
\label  {transfo}
 \Psi\rightarrow U \Psi U^\dagger,
\end{eqnarray}
and $\Psi$ is  postulated to transform  as indicated on the
right-hand side.

 That the equation, the transformation and symmetry operators $U$,
 and the solutions $\Psi$  occupy the same space is not only
 economical
but  it befittingly   implements  quantum mechanics, for it
ultimately implies measuring apparatuses are not constituted
differently in principle from the objects they measure.

  $U$ and $\Psi$ can be classified in terms of  Clifford
algebras. In four dimensions (4-$d$) $U$ is conventionally a
$4\times 4$ matrix
  containing symmetry operators as the Poincar\'e generators,
  but it  can contain others, although,     e. g., in the  chiral  massless case
it   can only carry an additional $U(2)$ scalar
symmetry$\cite{Jaime}$.
More symmetry operators appear
 if Eq. \ref{Jaimeq},  $\mu=0,...,3,$ is assumed within  the larger Clifford
algebra ${\mathcal C}_{N},$
 $\{ \gamma_\mu ,\gamma_\nu \} =2 g_{\mu\nu}$, $\mu,\nu=0,...,N-1$, where   $N$ is the
 (assumed even) dimension,
 whose structure is
  helpful in classifying the available symmetries, and which
  is represented  by  $2^{N/2}\times 2^{N/2}$
 matrices.
  The  usual 4-$d$ Lorentz symmetry,  generated in terms of
   $\sigma_{\mu\nu}=\frac{i}{2}[ \gamma_\mu,\gamma_\nu
 ],$  $\mu ,\nu =0,...,3$, is
 maintained and
  $U$  contains also $ \gamma_a$, $a=4,...,N-1$,  and their
 products as possible symmetry
 generators.
Indeed, these elements are  scalars for they commute
 with the Poincar\'e generators, which contain  $\sigma_{\mu\nu}$, and
they are also  symmetry operators  of  the massless  Eq.
\ref{Jaimeq},  bilinear in the $\gamma_\mu$ matrices, which is not
  necessarily the case for mass terms (containing  $\gamma_0$).
 In addition, their products with
   $\gamma_5=-i\gamma_0\gamma_1\gamma_2\gamma_3$  are Lorentz  pseudoscalars.
 As
 $[\gamma_5,\gamma_a]=0$,
  we can classify the (unitary) symmetry algebra as
  ${\mathcal S}_{N-4}={\mathcal S}_{(N-4)R} \times {\mathcal S}_{(N-4)L},$
  consisting of the projected
 right-handed ${\mathcal S}_{(N-4)R}=
\frac{1}{2}(1+\gamma_5)U(2^{(N-4)/2})$ and left-handed ${\mathcal
S}_{(N-4)L}= \frac{1}{2}(1-\gamma_5)U(2^{(N-4)/2})$
  components.

The solutions of Eq. \ref{Jaimeq} do not span all the matrix
 complex space, but this is achieved by considering also solutions of
\begin{eqnarray}
\label {Jaimeqnext} \Psi\gamma_0( -i \stackrel{\leftarrow}{
\partial_\mu}\gamma^\mu -M) ={ 0},
\end{eqnarray}
 consistent with the transformation in Eq. \ref{transfo},
 (the Dirac operator  transforming accordingly).

It is not possible to find always solutions that simultaneously
satisfy equations of the type   \ref{Jaimeq} and \ref{Jaimeqnext}
(except trivially), which means they are not simultaneously
on-shell, but they satisfy at least one and therefore the
Klein-Gordon
 equation. Indeed, the solutions of eqs. \ref{Jaimeq} and \ref{Jaimeqnext}, can
 be generally characterized as  bosonic  since
  $\Psi$ can be
understood to be formed of spinors as $\sum_{i,j} a_{ij} |w_i
\rangle\langle w_j |$.

Generalized operators acting on this  tensor-product space (spinor
$\times$ spinor $ \times$ configuration or momentum space) further
characterize the solutions. Positive-energy solutions, according
to Eq. \ref{Jaimeq} are interpreted as negative-energy solutions
from the right-hand side. This problem is overcome if we assume
the hole interpretation for the $\langle w_j |$ components, which
amounts to the requirement that operators generally acting from
the right-hand side acquire a minus, and that   the commutator be
used for operator evaluation.
 Thus, the 4-$d$ plane-wave solution
combination $\frac{1}{4}[(1- \gamma_5)\gamma_0(\gamma_1-i
\gamma_2)]e^{-i kx}$, with $k^\mu=(k,0,0,k),$ is a massless
vector$-$axial ($V-A$) state propagating along $\hat {\bf z}$ with
left-handed circular polarization, normalized covariantly
according to $\langle    \Psi_A
  | { \Psi_B}\rangle =tr\Psi_A^\dagger \Psi_B,$ the generalized inner
product for the solution space. In fact, combinations of
solutions of Eqs. \ref{Jaimeq} and \ref{Jaimeqnext} can be formed
with a well-defined
  Lorentz
index:   vector  $\gamma_0\gamma_\mu$, pseudo-vector
$\gamma_5\gamma_0\gamma_\mu$, scalar $\gamma_0$, pseudoscalar
$\gamma_0\gamma_5$, and antisymmetric tensor
$\gamma_0[\gamma_\mu,\gamma_\nu]$.
  For example,
 ${ A}^{\mathcal C}_\mu(x)
=\frac{i}{2}\gamma_0\gamma_\mu e^{-i    kx}$
 is a combination  that
 transforms  under parity into $ { A}^{{\mathcal C}\mu} (\tilde x ),$ $\tilde x_\mu=x^\mu,$ that is, as a vector.
 We may also view $\frac{1}{2}\gamma_0\gamma_\mu$ as  an orthonormal polarization
 basis, $A_\mu=tr\frac{1}{2}\gamma_\mu A^\nu\frac{1}{2}\gamma_\nu$\footnote{As for
 $\bar \psi=\psi^\dagger\gamma_0$, a unitary transformation can be
 applied to
 the fields and operators to convert them to a covariant form.}; just as
   $n_\mu$ in
  $A_\mu=g_{\mu\nu}A^\nu =n_\mu\cdot A^\nu n_\nu$.
In fact, the sum of  Eqs.  of \ref{Jaimeq} and
 \ref{Jaimeqnext} implies $\cite{Bargmann}$  for a $\Psi$ containing $\gamma_0\slash\!\!\!\! A={ A}^\mu\gamma_0\gamma_\mu $  that  ${
 A}^\mu$ satisfies the free Maxwell's equations.

Solutions   contain also products of $\gamma_a$ matrices
 that define their scalar-group representation.
 For given  $N$, there are variations of  the symmetry algebra
depending on the chosen Poincar\'e generators and Dirac equation,
respectively, through the  projection operators ${\mathcal P}_P,\
{\mathcal P}_D\in {\mathcal S}_{N-4}$, $[{\mathcal P}_P,{\mathcal
P}_D]=0$. ${\mathcal P}_P$ acts as in, e.g., ${\mathcal
P}_P\sigma_{\mu\nu}$, and
 ${\mathcal P}_D$ modifies  Eqs.
\ref{Jaimeq} and \ref{Jaimeqnext}  through ${\mathcal
P}_D\gamma_0( i
\partial_\mu\gamma^\mu -M).$ Together, they
characterize   the  Lorentz and scalar-group solution
representations.   We require     ${\rm rank}{\mathcal P}_D \le
  {\rm rank} {\mathcal P}_P$, for otherwise pieces of the solution
space exist that do not transform properly. For ${\mathcal P
}_D\neq 1$  Lorentz operators  act trivially on one side  of the
solutions containing $ 1-{\mathcal P }_P $,  since $(1-{\mathcal P
}_P){\mathcal P }_P=0$,  so we also get fermions.
  Fig. 1{(a)} depicts the
distribution of Lorentz-representation solutions according to the
matrix space they occupy in ${\mathcal S}_{6}$, when ${\mathcal
P}_P =
  {\mathcal P}_D \neq 1$. Although it
refers to $N=10$ case, it is  general for  any $N$.

An interactive field theory can be constructed in terms of the
above  degrees of freedom.
We consider a  vector  and fermion non-abelian gauge-invariant
theory. The expression for the kinetic component of the Lagrangian
density ${ \mathcal L}_{V}=
-\frac{1}{4}F_{\mu\lambda}^ag^{\lambda\eta}\delta_{ab} F^{b\mu}_{\
\  \eta}=-\frac{1}{4 N_o}tr {\mathcal   P}_D
F_{\mu\lambda}^a\gamma_0\gamma^{\lambda} G_a F^{b\mu}_{\ \
\eta}\gamma_0\gamma_ \eta G_b$
 shows ${\mathcal   L}_{V}$
is equivalent to a trace over combinations over normalized
components $\frac{1}{\sqrt{ N_o}}\gamma_0\gamma_\mu G_a
$
with coefficients $F^a_{\mu\nu}=\partial_\mu A_\nu^a-
\partial_\nu^a A_\mu^a+ g A_\mu^b A_\nu^c C^a_{bc}$, $g$ the
coupling constant,  $\gamma_\mu\in{\mathcal C}_{N}$, $G_a\in
{\mathcal S}_{N-4} $ the group generators, $C^a_{bc}$ the
structure constants,
  and $N_o=trG_a G_a $, where for
non-abelian irreducible representations we use $trG_i G_j=2
\delta_{ij}$.

  Similarly,   the interactive part of the fermion gauge-invariant Lagrangian
$ {\mathcal   L}_{f}=\frac{1}{2}{\psi^\alpha}^\dagger\gamma_0(
i\stackrel{\leftrightarrow}\partial_\mu-g A^a_\mu G_a
)\gamma^\mu\psi^\alpha,$
 with $\psi^\alpha$  a  massless spinor with flavor $\alpha$,
can be written
   ${\mathcal   L}_{int} = -g\frac{1}{ 2 N_o} tr
{\mathcal P}_D A_{\mu}^a \gamma_0\gamma^\mu G_a
j^{b\alpha}_\lambda \gamma_0\gamma_\lambda G_b,$    with  $
j_\mu^{a\alpha}=tr{\Psi^\alpha}^\dagger\gamma_0\gamma_\mu G_a
\Psi^\alpha$
 containing $\Psi^\alpha=\psi^\alpha\langle \alpha |$, and $\langle \alpha|$
 is a row  state accounting for the flavor.
 ${\mathcal   L}_{int}$ is written in terms of  $\gamma_0\slash\!\!\!\! A$,
 and $\gamma_0\slash\!\!\!j^{a\alpha}$,
 that is, the vector field and the current occupy   the same spin space. This
connection and the quantum field theory (QFT) understanding of
this vertex as the transition operator between fermion states,
exerted by a vector particle, with the coupling constant as a
measure of the
 transition probability,  justifies the  interpretation  for it $\frac{1}{2}g A^{a\mu} j^{a\alpha}_\mu= A^{a\mu}\frac{1}{\sqrt{ N_o}}
tr{\Psi^\alpha}^\dagger \gamma_0\gamma_\mu G_a\Psi^\alpha$,
leading to the  identification $g\rightarrow 2\sqrt{ \frac{K}{
N_o}}$, $K$ correcting for  over-counted  reducible
representations, which is further clarified below. The theoretical
assignment of $g$ complements QFT, in which the coupling constant
 is set
 experimentally.
  It should be also
understood as tree-level information, while the values  are
modified by the presence of a virtual cloud of fields, at given
energy. Although in QFT the coupling constant
 is obtained  perturbatively in terms of powers of
 the bare, which takes  infinite values  absorbed through renormalization,
 we may take the view that renormalization is a
calculational device and that its physical value  is a
manifestation of the bare one;
 this is feasible for  small  coupling constants,  which  can give
 small corrections. Energy corrections are also necessary for a more detailed
calculation.

As for the initial formulation, ${\mathcal P}_D$ restricts the
possible gauge symmetries that can be constructed in the
Lagrangian, for $ \gamma_0\gamma_\mu G_i$ needs to be contained in
 the space it projects. Thus,
${\mathcal P}_P$ and ${\mathcal P}_D$ determine the symmetries,
   which are global, and  in turn,
determine the  allowed gauge interactions. Furthermore, they
 fix the representations, assumed physical.
The $N=6$ case has been
 researched$\cite{Jaime}$ and connections have been found to the
 $SU(2)_L\times U(1)$, electroweak sector of the SM.
 It is   apparent that the minimal algebra that includes the
SM groups   requires $N=9+1$, with  ${\mathcal S}_6$, on which we
will concentrate.
 There are  limited ways in which we may represent the SM
 interactions in such a matrix space and only one giving the correct fermion
 and boson quantum numbers.  In order to have
fermions we need ${\mathcal P}_P \neq 1$. Account of the quark
quantum numbers requires that their left-handed $SU(3)\times
 SU(2)_L$, and right-handed   $SU(3)\times
 U(1)_Y$ symmetry generators   be direct-product reducible representations
 occupying,   respectively,   $6\times 6$ matrix pieces of ${\mathcal S}_{6L}$
 and   ${\mathcal S}_{6R}$. The remaining  $2\times 2$
 matrix   into which  ${\mathcal S}_{6L}$
is broken is associated  to the $SU(2)_L$ acting on the
$SU(3)$-singlet
 leptons, and that of ${\mathcal S}_{6R}$ to
a $U(1)$  describing the right-handed leptons hypercharge, and an
inert $U(1)$ that gives  rise to two fermion
 generations (all this applies also to antiparticles).
  There are additional $U(1)$ symmetries which can be  assigned in
correspondence to SM symmetries.

To represent  these ${\mathcal S}_{6}$ terms, we use the 64
matrices composed of 6  $\tilde\gamma_{a-4}=\gamma_{a}$,
$a=5,...,9$, $\tilde \gamma_0=i \gamma_4$, 15 pairs
$\tilde\gamma_{ab}=\tilde\gamma_a\tilde \gamma_b$, $a< b,$ etc.,
20 triplets $\tilde\gamma_{abc}=\tilde \gamma_{a} \tilde
\gamma_b\tilde \gamma_{c}$, 1 sextuplet
$\bar\gamma_7=\tilde\gamma_0\tilde\gamma_1\tilde\gamma_2\tilde\gamma_3\tilde\gamma_4\tilde\gamma_5,$
 15 quadruplets
$\tilde\gamma_{7ab}=\tilde\gamma_7\tilde\gamma_{ab}$, 6
quintuplets  $\tilde \gamma_{7a}=\bar\gamma_7\tilde\gamma_a$,
 and the identity $1$.

 ${\mathcal S}_{6}$ has  a Cartan algebra of dimension 8+8,
 which  we write in terms
 of the projection operators $P_{R}=\frac{1}{4}(1-\tilde\gamma_0)
 (1+\tilde\gamma_{73})$, $P_{S\pm}= \frac{1}{2}(1\mp i\tilde\gamma_{26})$. The
 latter
 contains
 the  baryon  numbers $P_{B(V\pm
A)}=\frac{1}{2}(1\pm\gamma_5)(1-P_{R})$, with  $V\pm A$
 components;   $SU(3)$ $\lambda_{3(V\pm
A)}=\frac{1}{2}P_{B(V\pm
A)}[P_{S+}(\tilde\gamma_4+\tilde\gamma_{73}) +
P_{S-}(\tilde\gamma_4+\tilde \gamma_{703})] $, $\lambda_{8(V\pm
A)}=\frac{1}{2\sqrt{3}}P_{B(V\pm A)}[P_{S+}
(-\tilde\gamma_4+\tilde\gamma_{73}-2  \tilde\gamma_{703})+
P_{S-}(-\tilde\gamma_4 -2\tilde\gamma_{73}+\tilde\gamma_{703})] $;
lepton numbers $P_{L(V+A)}=\frac{1}{2}(1+\gamma_5)P_{R}P_{S+},$
$P_{L(V-A)}=\frac{1}{2}(1-\gamma_5)P_{R};$
 $SU(2)_L$
$I_3= (P_{B(V-A)}+P_{L(V-A)})i\tilde\gamma_{26};$  flavor
operators generated by $P_{F}=\frac{1}{2}(1+\gamma_5)P_{R}P_{S-}$
(and additional); hypercharges $Y_\pm=
\frac{1}{2}(1\pm\gamma_5)[P_{L(V-A)}- \frac{1}{3}P_{B(V-A)}+2
P_{L(V+A)}+ \frac{1}{3}P_{B(V+A)}(2 P_{S+}-4P_{S-})]$; isocolor
$\bar\lambda_{j(V- A)}=\lambda_{j(V- A)}I_3$,  and hypercolor
$\bar\lambda_{j(V+A)}=\frac{1}{3}\lambda_{j(V+A)}
 P_{B(V+A)+}(4
P_{S+}+2P_{S-})$, $j=3,8$.

The  choice $Y=Y_++Y_-$  that gives the correct hypercharge
fermion quantum numbers can be deduced from the condition that it
not be axial, separately for quarks and leptons, namely,
$tr\gamma_5  YL  =0,$ and $tr\gamma_5 YB =0$, where $L=P_{L(V+A)}+
P_{L(V-A)},$ $B=P_{B(V+A)}+ P_{B(V-A)}$,  which leads to the
correct ratios between right-handed and left-handed lepton and
quark hypercharges. The anomaly-cancellation condition sets the
ratio between lepton and quark hypercharges. For $N=6$,
$Q=\frac{1}{2}Y+I_3 $ can also be deduced$\cite{Jaime}$ as one of
the operators commuting with the Hamiltonian when mass terms are
added, which sets the form of $Y$.

 ${\mathcal P}_P={\mathcal P}_D= B+L\in  {\mathcal S}_{6}$ is
the only option describing SM fermions. In Fig. 1{(a)}-{(d)}
are the solution representations resulting from such massless
Hamiltonian, classified with the above generators.
 Each of the four types of solution
 in  {(a)} is an $8\times 8$ matrix, three of which are
given  explicitly in {(b)}-{(d)}
with the solutions' quantum numbers. {(b)} and {(d)} contain
 vectors (and axial-)   $\gamma$-bilinear  solutions  that  are in the
adjoint representation, with {(b)} also containing the flavor
group, and fermion solutions. {(c)} has $\gamma$-linear solutions
conformed of scalar (and pseudo-), antisymmetric tensors, and
fermions.  Fixing ${\mathcal P}_D$, the   physically feasible
($trG_i G_j=2 \delta_{ij}$) gauge groups are
comprised$\cite{Weinbergbook}$ of the $U(1)$ and compact simple
algebras generated in it.  With the conditions of
anomaly-cancelling, renormalizability, $B,$  $L$ conservation, the
allowed Lagrangian reduces basically to that of the SM, with
particles with correct Lorentz and gauge group representations:
gluons,  weak,  and hypercharge vectors.
 A Higgs-like particle can also be constructed from the
neutral and charged scalar   $SU(2)_L$  doublets in  {(c)}. There
are two generations of $SU(2)_L$ lepton and quark doublets
$(e,\nu)_L,$ $(u,d)_L$  and $e_R,$ $u_R,$ $d_R$ singlets
(generation choice arbitrary). We also find particles beyond the
SM as   scalar leptoquarks in {(b)},  and $V-A$ iso-gluons in
{(c)}.

The
  vector-field normalized polarizations  provide the coupling
constants.  The physics guides in obtaining  the field
configuration, pointing at
 overcounted degrees of
freedom in $K$ direct-product reducible representations, and
gives a clue on the energy scale. For the hypercharge
$B_\mu=\frac{g^\prime}{2} Y\gamma_0\gamma_\mu$, so $g^\prime=2/[{2
(2 + 2^2 + 6(\frac{1}{3})^2+   3 (\frac{2}{3})^2 +3
(\frac{4}{3})^2)]^{1/2}}$ $=\frac{1}{2}\sqrt{\frac{3}{5}}\approx
.387$.
 The normalized
weak-boson   component $W_{B\mu}^{({\bf 3},{\bf 1})}$ (in direct
product with $SU(3)$ space) needs to be contracted to $\bar
W_{B\mu}^{({\bf 3},{\bf 1})}$, leading to  its coupling to be
rescaled by $\sqrt{3},$ which gives the same couplings to quarks
as to leptons (see   {(d)}).
 Gluons  are described by $\frac{1}{\sqrt{2}}(A^{({\bf
1},{\bf 8})}_\mu+\tilde A^{({\bf 1},{\bf 8})}_\mu)$ terms (see
 {(b),}   {(d)}.) One $2^{1/2}$ factor  corrects each
$V\pm A$  components (same coupling to chiral and massive quarks)
 and another $2^{1/2}$ corrects for the $SU(2)_L$ and hypercharge
products. This gives $ g_s =2^2 /[{2 (2^3 )]^{1/2}}=1,$ or
$\alpha_s=\frac{g_s^2}{4 \pi}\approx .080.$ In the case of the
physical weak field $W^{({\bf 3},{\bf 1})}_\mu=
\frac{1}{\sqrt{2}}(\bar W_{L\mu}^{({\bf 3},{\bf
1})}+W_{B\mu}^{({\bf 3},{\bf 1})})$,
 we assume a single $SU(2)_L$ irreducible representation, for $W_\mu^{({\bf 3},{\bf 1})}$ acts
separately on quarks and leptons.
 Then  $g =
2 /[{2 (2^2)]^{1/2}}=\frac{1}{\sqrt{2}}\approx .707$.
 From the  demand that  a vector particle be obtained with  the correct
parity, or others$\cite{Jaime}$, $Q$ is derived  as the only
non-trivial scalar commuting the
 Hamiltonian, leading to an expression  for
the photon $A_\mu=\frac{1}{\sqrt{g^2+{g^\prime}^2}}(g
 B_\mu+g^\prime W^0_\mu)$.
 One gets Weinberg's angle $\theta_W$ from here.
It can also
 be consistently obtained    directly from $g$ and $g^\prime$.
  From the SM\cite{Glashow}
$tan(\theta_W)=g^\prime/g$,
 so $sin^2(\theta_W)=3/13\approx .23078.$
 The assumed fermion massive terms and
electroweak symmetry-breaking conditions suggest a
     comparison of  these numbers with experimental values
at energies of order $M_Z.$ These are$\cite{tables}$, one
standard-deviation last-ciphers uncertainty in parenthesis,
$g_{ex}^\prime=.35743(8)$, $sin^2(\theta_{Wex})= .23117(16),$ and
$\alpha_{s(ex)}=.1185(20)$.    Unified-generator  fields are
obtained when assuming that $Y$ and $I_3$ belong to the same
group, thus having the same normalization convention, so
$g^\prime$ needs to be rescaled to
$g_1^{uni}=\sqrt{\frac{5}{3}}{{g^\prime}^{uni}}$. From the
assumption that $g_1=g$
 and  that  $I_3$ acts equally on  massless chiral  leptons or quarks, we
get $g^{uni}=2/{[2(8)]^{1/2}}=\frac{1}{2}$ ($K=1$), and recover
the $SU(5)$ unification result$\cite{quinn}$
$sin^2(\theta_W^{uni})=3/8$.

 The theory thus presented succeeds in
reproducing many aspects of the  SM. We obtain the groups
$SU(2)_L\times SU(3)\times U(1)$, with corresponding  vector
bosons  acting on quarks  and leptons in two generations with
correct quantum numbers and chiralities. A Higgs particle is also
obtained. By using the allowed vertices with
normalized-polarisation fields we are able to calculate coupling
constants, around electroweak breaking. It is not obvious that SM
features can be described and yet a special model configuration
among few possible gives such predictions. An assortment  of new
particles beyond the SM are also obtained but not all need be
stable or  appear at low energies.

With  the  Poincar\'e  and SM-gauge symmetric  Lagrangian
presentation of the model, renormalization and quantization can be
applied, leading to a QFT formulation. Its simplicity, with   spin
as its basic building block, should allow for a generalization and
application in theories such as supersymmetry$\cite{wess}$ and
those accounting for gravity, such as supergravity$\cite{wess}$
and strings.

There is a robustness to the  predictions, for extension of the
model into  such structures,  requiring  additional reducible
representations and  quantum numbers,
 will not change the results.
 On the other hand,   inclusion into other models with different irreducible
representations should modify them, making  the $N=9+1$ case
unique.

 The close connection between
the results hence derived and the physical particles'
phenomenology makes plausible the idea that, as with the spin,
  the gauge vector and matter fields, and their
interactions  originate in the structure of space-time.

\setcounter{equation}{0}


 The
author acknowledges support from  DGAPA for project IN118600   at
the UNAM.

\def\baselinestretch{.9}


\def\baselinestretch{1.1}
\renewcommand{\arraystretch}{.85}

\begin{table}[h]
\center{ \begin{tabular}{||c|c||c ||} \hline\hline
 & \raisebox{-1.0ex}{ F } & \raisebox{-1.0ex}{ F } \\  [0.2 cm]
\hline
&    &  \\
   &    &  \\
   &    &  \\
 \raisebox{-0.2ex}{F}  & \raisebox{-0.2ex}{V} & \raisebox{-0.2ex}{S,A}    \\ 
 &    &  \\
 &    &  \\
   &  \ \ \ \ \ \ \ \  \ \ \  { \small {{(b)}}}  & \ \ \ \ \ \ \ \ \ \ \ \ \ \  \ \ \ { \small {{(c)}}} \\
\hline\hline
 &    &  \\
 &    &  \\
 &    &  \\
  \raisebox{-1.5ex}{F}  & \ \ \ \ \  \raisebox{-1.5ex}{S,A}  \ \ \ \ \ &   \ \ \ \ \ \  \ \ \  \raisebox{-1.5ex}{V} \ \ \ \ \ \  \ \ \ \ \  \\
  &    &  \\ 
  &    &  \\
   &    &  \\
   &    & \ \ \ \ \ \ \ \ \ \ \ \ \ \ \  \ \ \ { \small {{(d)}}} \\
\hline\hline
\end{tabular}}\\[.5cm] {{(a)} }\\
\label{tab:tablejbFBV} {Arrangement of ${\mathcal S}_6$ scalar
components of $N=9+1$ solutions. { (a)} ${\mathcal S}_6$ is
divided into four 6-$d$ $8\times 8$ matrix blocks, with fermion
(F), vector (and axial-) (V), and scalar (and pseudo-) and
antisymmetric  (S,A) terms. }
\end{table}
\renewcommand{\arraystretch}{1}
\normalsize

\ \

\

\

\

\newpage

\def\baselinestretch{1.1}
 \parskip 5pt plus 1pt

\renewcommand{\arraystretch}{1.45}
\begin{table}[h]
\center{ \begin{tabular}{|c |c|c | c|c|} \hline \   \   \  \
\hspace{-.01cm} & $ {\Large e}^2_L \hspace{-.2cm}  $  &  $\bar u_{
L}^{-4/3({\bf 1},{\bf \bar 3})} $
 & ${\bar {d}_{ L}}^{\ 2/3({\bf 1},{\bf \bar 3})} $\\ \hline &  \hspace{-.25cm} $B_{\mu}^{(-2)} $ \hspace{-.35cm}  &
$A_\mu^{-10/3({\bf 1},{\bf \bar 3})
 }$
 & $  A_\mu^{-4/3({\bf 1},{\bf
\bar 3}) }  $\\
\hline &
 &  \hspace{-.3cm}    \raisebox{-1.6ex}{ $  B_{\mu}^{(4/3)}    \  \tilde A_{\mu }^{({\bf 1},{\bf 8})}
 \
 $ } \hspace{-.4cm}  &    \hspace{-.3cm}  \\
  &   &   &  $ \ \ B_{\mu}^{ 2} \ \   A_\mu^{2({\bf 1},{\bf 8})} $     \\
 &
 &  \raisebox{1.6ex}{ \ $  \bar  A_{\mu }^{({\bf 1},{\bf 8})}    \
 $ }&     \\
\hline &
   &  &  \hspace{-.2cm}
\raisebox{-1.6ex}{  $  B_{\mu}^{(-2/3)}  \ \tilde A_\mu^{({\bf
1},{\bf 8})}
  $ }  \hspace{-.4cm}
  \\
  &   &   &  \\
   &
   &  &
\raisebox{1.6ex}{  $  \bar  A_\mu^{({\bf 1},{\bf 8})}
  $ }
  \\
\hline
\end{tabular}}\\[.5cm] { {(b)}}\\
\label{tab:tablejbVF}{(b) } $\frac{1}{2}
(1+\gamma_5)\gamma_0\gamma_\mu$ fields with (throughout)
notation  $X^{Y{\bf(i,c) }}$, where $Y$ is the hypercharge,
omitted for $Y=0$, except for  fields $X^{(Y) },$ labelled
according to the particle's $Y$ they give, and ${\bf i}$, ${\bf
c}$ label respectively the $SU(2)_L$ isospin, $SU(3)$ color
representations, omitted for singlets ${\bf(1,1) }$. Here and on,
all empty places are occupied by antiparticles, except for the
flavor-assigned upper-left box here. We get $V+A$ hypercharge
carriers $B_\mu^{(-2)},$ $B_\mu^{(-4/3)},$ $B_\mu^{(2/3)},$ gluons
$ \tilde A_{\mu}^{({\bf 1},{\bf 8})}$, hyper-gluons $\bar
A_\mu^{({\bf 1},{\bf 8})}$, $ A_\mu^{2({\bf 1},{\bf 8})}$;
hyper-triplets $\tilde A_\mu^{-10/3({\bf 3},{\bf \bar 3})}$,
$A_\mu^{-4/3({\bf 1},{\bf \bar 3})}$; and isospin-singlet leptons
${\Large e_L},$ and quarks   $\bar u_{ L} $
 $\bar d_{ L} $.
\end{table}
\renewcommand{\arraystretch}{1}
\normalsize
\newpage

\def\baselinestretch{1.1}
 \parskip 5pt plus 1pt

\renewcommand{\arraystretch}{1.45}
\begin{table}[h]
\center{ \begin{tabular}{|c|c |c|} \hline  $ {\large e}_R^{1({\bf
2},{\bf 1})}         {\large {\bar  \nu }}_R^{1({\bf 2},{\bf 1})}
$ \hspace{-.3cm} & \multicolumn{2}{c|}{\ \ \ $\bar u_{
R}^{-1/3({\bf 2},{\bf \bar 3})} \ \ \
    \ \bar  d_{ R}^{\ -1/3({\bf 2},{\bf \bar 3})}$  \  \  \  \ }  \\
  \hline
 $\phi^{-1({\bf 2},{\bf 1})}_l$ &
 \multicolumn{2}{c|}{$\phi^{-7/3({\bf 2},{\bf \bar 3})}$}
\\ \hline
    &
  \multicolumn{2}{c|}{ \ \ }   \\
  $\phi^{7/3({\bf 2},{\bf  3})}$   &
  \multicolumn{2}{c|}{$\phi^{1({\bf 2},{\bf  1})}_q\ \ \ \ \ \ \phi^{1({\bf 2},{\bf  8})}$}   \\
\ \    &
  \multicolumn{2}{c|}{ \ \  }   \\
\hline &
  \multicolumn{2}{c|}{ \ \ }   \\
 $\phi^{1/3({\bf 2},{\bf  3})}$
   &  \multicolumn{2}{c|}{$\phi^{-1({\bf 2},{\bf  1})}_q\ \ \ \ \ \ \ \phi^{-1({\bf 2},{\bf  8})}$}
  \\
  &
  \multicolumn{2}{c|}{ \ \ }   \\
\hline
\end{tabular}}\\[.5cm] {{(c)} }\\
\label{tab:tablejbBF}{(c) } $\frac{1}{2}
(1+\gamma_5)\gamma_0\gamma_\mu\gamma_\nu$ fields. We find
Higgs-like scalars   $\phi^{-1({\bf 2},{\bf 1})}_l$ acting on
leptons, and $\phi^{1({\bf 2},{\bf 1})}_q,$ $\phi^{-1({\bf
2},{\bf 1})}_q$ on quarks; leptoquarks $\phi^{ 7/3({\bf 2},{\bf
3})}$, $\phi^{  -7/3({\bf 2},{\bf \bar 3})}$, $\phi^{1/3({\bf
2},{\bf 3})}$; iso-octets $\phi^{1({\bf 2},{\bf 8})}$,
$\phi^{-1({\bf 2},{\bf 8})}$;  and isospin-doublet leptons
$(e,\bar \nu)_R$ and quarks $(\bar u,\bar d)_R $.
\end{table}
\renewcommand{\arraystretch}{1}
\normalsize
\newpage

\def\baselinestretch{1.1}
 \parskip 5pt plus 1pt

\renewcommand{\arraystretch}{1.45}
\begin{table}[h]
\center{ \begin{tabular}{|c@{\hspace{0cm}}
|c@{\hspace{0cm}}|c@{\hspace{0cm}} | c|} \hline
 \raisebox{-3.5ex}{ \hspace{-.33cm} $B_{\mu}^{(-1)}    \hspace{ .05cm}   W^{({\bf 3},{\bf 1})}_{L\mu}$ } \hspace{-.25cm}   &
 \multicolumn{2}{c|}{\raisebox{-3.5ex}{ $B_{ \mu}^{-4/3({\bf 1},{\bf 3})} \ \ \  A_\mu^{-4/3({\bf 3},{\bf 3})}$}}
 \\ [-0.12 cm]
  &   \multicolumn{2}{c|}{\ }  \\ [-0.12 cm] \hline
 & \raisebox{-1.6ex}{    $  B_{\mu}^{(1/3)}  \ W^{({\bf 3},{\bf 1})}_{B\mu}
 $ }&
 \\  &   &  $W^{({\bf 3},{\bf 1})}_{B\mu}   \     A_\mu^{({\bf 3},{\bf 8})}$     \\
  & \raisebox{1.6ex}{     $A_\mu^{({\bf 1},{\bf 8})}    \  A_\mu^{({\bf 3},{\bf 8})}$     \ } &   \\
\hline
   &  &
 \raisebox{-1.6ex}{     $  B_{\mu}^{(1/3)}  \   W^{({\bf 3},{\bf 1})}_{B\mu}
 \
  $ }
  \\
   &   &  \\
& &\raisebox{1.6ex}{      $A_\mu^{({\bf 1},{\bf 8})}    \  A_\mu^{({\bf 3},{\bf 8})}$ \  } \\
 \hline
\end{tabular}}\\[.5cm] { {(d)}}\\
\label{tab:tablejbV}{(d)}  $V-A$ terms with $\frac{1}{2}
(1-\gamma_5)\gamma_0\gamma_\mu$  form, consisting of weak vectors
$W_{B\mu}^{({\bf 3},{\bf 1})}$ acting on quarks, and
$W_{L\mu}^{({\bf 3},{\bf 1})}$ on leptons; gluons $A_\mu^{({\bf
1},{\bf 8})}$, iso-gluons $A_\mu^{({\bf 3},{\bf 8})}$;
hypercharges $ B_{\mu}^{(-1)}$ for leptons,  $ B_{\mu}^{(1/3)}$
for quarks; hypertriplets $B_\mu^{-4/3({\bf 1},{\bf 3})},$ and
isotriplets $A_\mu^{-4/3({\bf 3},{\bf 3})}$.
\end{table}
\renewcommand{\arraystretch}{1}
\normalsize

\end{document}